\documentclass[iop]{emulateapj}
\pdfoutput=1
\usepackage{graphicx,natbib}
\usepackage{subfigure}
\usepackage{apjfonts}
\usepackage[colorlinks=true,linkcolor=blue,citecolor=blue]{hyperref}
   
\slugcomment{Submitted to ApJL}

\newcommand{\tmerge}{$t_{\rm m}$}
\newcommand{\porb} {$P_{\rm orb}$}
\newcommand{\msun} {$M_{\odot}$}
\newcommand{\mone} {$M_{1}$}
\newcommand{\mtwo} {$M_{2}$}
\newcommand{\rsun} {$R_{\odot}$}
\newcommand{\rtwo}{$R_2$}
\newcommand{\rlone}{$R_{\rm L1}$}
\newcommand{\lsun} {$L_{\odot}$}

\newcommand{\beqn}{\begin{equation}}
\newcommand{\eeqn}{\end{equation}}

\def\Fref#1{Figure~\ref{fig:#1}}
\def\Sref#1{Section~\ref{sec:#1}}

\begin{document}

\shortauthors{Macias et al.}
\shorttitle{The Past and Future of DWDs}
\title{The Past and Future of Detached Double White Dwarfs with Helium Donors}

\author{Phillip J. Macias, Monique Windju, and Enrico Ramirez-Ruiz}
\affiliation{Department of Astronomy and
  Astrophysics, University of California, Santa Cruz, CA
  95064}
   
\begin{abstract} 
We present a method for modeling the evolution of detached double white dwarf (DWD)  binaries hosting helium donors  from the end of the common envelope (CE) phase to the onset of Roche Lobe overflow (RLOF).  This is achieved by  combining detailed stellar evolution calculations of extremely low mass (ELM) helium  WDs possessing hydrogen  envelopes with the  the orbital shrinking of the binary driven by gravitational  radiation. We show that the consideration of hydrogen fusion in these systems is crucial, as a significant fraction ($\approx$50\%)  of future  donors are expected to still  be burning  when  mass transfer commences. We apply our method to two detached eclipsing DWD systems, SDSS J0651+2844 and NLTT-11748, in order to demonstrate the effect that carbon-nitrogen-oxygen (CNO) flashes have on constraining the evolutionary history of such  systems. We find that when CNO flashes are absent on the low mass WD  ($M_{2} \lesssim$ 0.18 \msun), such as  in NLTT-11748,  we are able to self consistently solve for the donor conditions at CE detachment given a reliable cooling age from the massive WD companion.    When CNO flashes occur (0.18 \msun $\lesssim$ $M_{2} \lesssim$ 0.36 \msun), such as  in SDSS J0651+2844, the evolutionary history is eradicated and we are unable to comment on the detachment conditions. We find that for any donor mass our models are able to predict the conditions at reattachment and comment on the stabilizing  effects of hydrogen envelopes. This method can be applied to a population of detached DWDs with measured donor radii and masses. 
\end{abstract}
\subjectheadings{binaries: close--- Galaxy: stellar content--- stars: general--- white dwarfs }

\section{Introduction}

Detached DWD binaries are  considered to be the progenitors of a variety of intriguing systems and explosive phenomena such as AM CVn stars, type .Ia and possibly type Ia supernovae \citep{Kilic2014b,Bildsten2007,Iben1984,Webbink1984,Guillochon2010,Pakmor2012}. 
Recently, considerable effort has been put into studying the stability of these systems when they come into mass transfer, as those which become dynamically unstable and merge have emerged as a viable candidate for type Ia supernovae throughout much of the \mone-\mtwo\, parameter space \citep{Dan2012}. 

Systems which will come into contact within a Hubble time (referred to as {\it merger systems}) are the end products of a violent and hydrodynamically rich past as they must have undergone one or two phases of unstable mass transfer driven by radius expansion during the giant phase \citep{Marsh1995}. During this phase, the ensuing single envelope embedding the two stars  brings the system to smaller separations. 
The details of this CE phase are a subject of active investigation as they are difficult to constrain both observationally and theoretically  \citep{Ivanova2013}. After the CE has ceased, retention of a small amount ($10^{-3}-10^{-2}$ \msun) of hydrogen on the donor will result in the ELM WD burning hydrogen in a thin shell and remaining bright for billions of years \citep{Althaus2001,Serenelli2001,Panei2007}.

As the systems are drawn closer together via gravitational radiation, they will eventually come into a phase of mass transfer. The system either merges or  survives to become an interacting binary. For  the system to survive the mass transfer must be stable, and this depends sensitively  on the mass ratio (defined as $q =$ \mtwo/\mone\, where \mone $>$ \mtwo\,  is the accretor mass), angular momentum  transport  mechanisms
and the donor's internal structure  \citep{Marsh2004,Gokhale2007,Dan2011,Kremer2015}.  

The discovery of tens of detached DWD systems with ELM donors (\mtwo $\lesssim$ 0.2 \msun)  has sparked an interest in the modeling of helium WDs with hydrogen envelopes, as their cooling ages depend on the details of the quiescent burning, the presence and effect of CNO induced thermonuclear flashes, and the insulating influence of their envelopes \citep{Althaus2013,Istrate2014b}. Of these detached DWDs, 25 are merger systems, including 5 with merging times $\lesssim$100 Myr \citep{Kilic2012,Kilic2014}. We demonstrate in this {\it Letter}  that during this pre-merger period we are able to make predictions about the thermal state of the donor at the beginning of RLOF, as well in some cases {\it retrodictions} about the conditions at the end of the difficult to constrain CE phase. We use two detached eclipsing DWD systems (SDSS J0651+2844 and NLTT-11748) as testbeds with a clear emphasis on the applicability of this method to others. In \Sref{hydrogen} we discuss the importance and prevalence of quiescent hydrogen burning. In \Sref{evolution} we present our method and apply it to  determining the future and past evolution of  SDSS J0651 and NLTT-11748.  We discuss our results and conclude in  \Sref{discussion}.

\section{The Role and Ubiquity of Hydrogen in DWDs}
\label{sec:hydrogen}
\citet{D'Antona2006} first explored the effect of hydrogen burning on an ELM WD donor at the onset of mass transfer with the aim of explaining the period decrease of the shortest orbital period binary, HM Cancri. The entropy injection due to the burning at the hydrogen/helium interface causes the WD radius to grow, reaching 10-100 times its zero temperature value depending on the amount of hydrogen in the envelope. As a result,  the WD  overflows its Roche lobe at larger separations. A key aspect is that the donor deviates from the standard WD mass-radius relation $R_2 \propto M^{-1/3}_2$ and it contracts as its hydrogen envelope is stripped. This has a two-fold effect: it reduces the equilibrium mass transfer rate and causes the period derivative to be dominated by gravitational wave losses as opposed to mass transfer. In this way, the presence of hydrogen serves to temporarily stabilize the binary regardless of the initial mass ratio.  

In order to determine the role hydrogen plays in the future dynamical evolution of the observed DWD merger systems, one can make  a simple  analytical estimate as to whether or not the donor WD will be quiescently burning at the onset of RLOF. Using the core mass - luminosity relation \citep{Paczynski1970}, \citet{Istrate2014b} derive a timescale for hydrogen burning 
\beqn
t_{\rm p} = 400 \hspace{1 mm} {\rm Myr} \left(\frac{0.20 M_{\odot}}{M_{\rm WD}}\right)^7,
\eeqn
assuming, for simplicity,  that all helium donor WDs have 0.01 \msun\, of hydrogen mass available to  burn in their envelopes. 

One can then  compare $t_{\rm p}$  with the merging timescale as derived using the binary's orbital parameters,
\beqn
t_{\rm m} = 9.838 \times 10^{6} \hspace{1 mm}{\rm yr} \left(\frac{P_{\rm orb}}{\rm 1 \hspace{1 mm}hr}\right)^{8/3} \frac{(M_1 + M_2)}{M_1 M_2}^{1/3},
\eeqn 
where \mone\, and \mtwo\, are measured in units of \msun. 

 \Fref{m1m2} shows the observed detached DWD population in the  \mone-\mtwo \, plane and the ratio $t_{\rm p} / t_{\rm m}$ given these simple assumptions. We find that a significant fraction (about 50\%) of the systems observed today may contain donor stars which will still be burning hydrogen at the onset of RLO. The extended envelope of the donor may render it more susceptible to tidal feedback into the orbit and also will cause it to begin mass transferring at a larger separation. 
    
These results point to the existence of a population of short period, accreting DWDs undergoing a period of stable mass transfer in which the period derivative is consistent with that expected from gravitational waves and a mass transfer evolution which is drastically different  from that  expected when the  cold He WD donor solution is applied. This is particularly important for systems in which mass transfer is expected to be dynamically unstable.  The two shortest orbital period binaries, HM Cancri and V407 Vulpeculae, exhibit this behavior \citep{Marsh2002, Israel2004, Barros2007}, indicating that the duration of this unique phase of mass transfer is certainly non-negligible. 

The presence of hydrogen on the donor at the onset of mass transfer is clearly crucial to the stability of DWDs, which is  in turn  sensitive to the hydrogen burning lifetime after the CE phase $t_{\rm p}$, which has been derived here under very simplistic assumptions. In reality, the evolution of ELM He WDs with hydrogen burning envelopes is complicated by the effects of diffusion, gravitational settling, and other mixing processes . Additionally (and perhaps more importantly), some WD core masses encounter a thermal instability in the hydrogen burning layer, resulting in hydrogen shell flashes which are difficult to describe analytically. These complications motivate us to model the evolution of these objects numerically.

  \begin{figure}
  \centering
\includegraphics[scale=0.82]{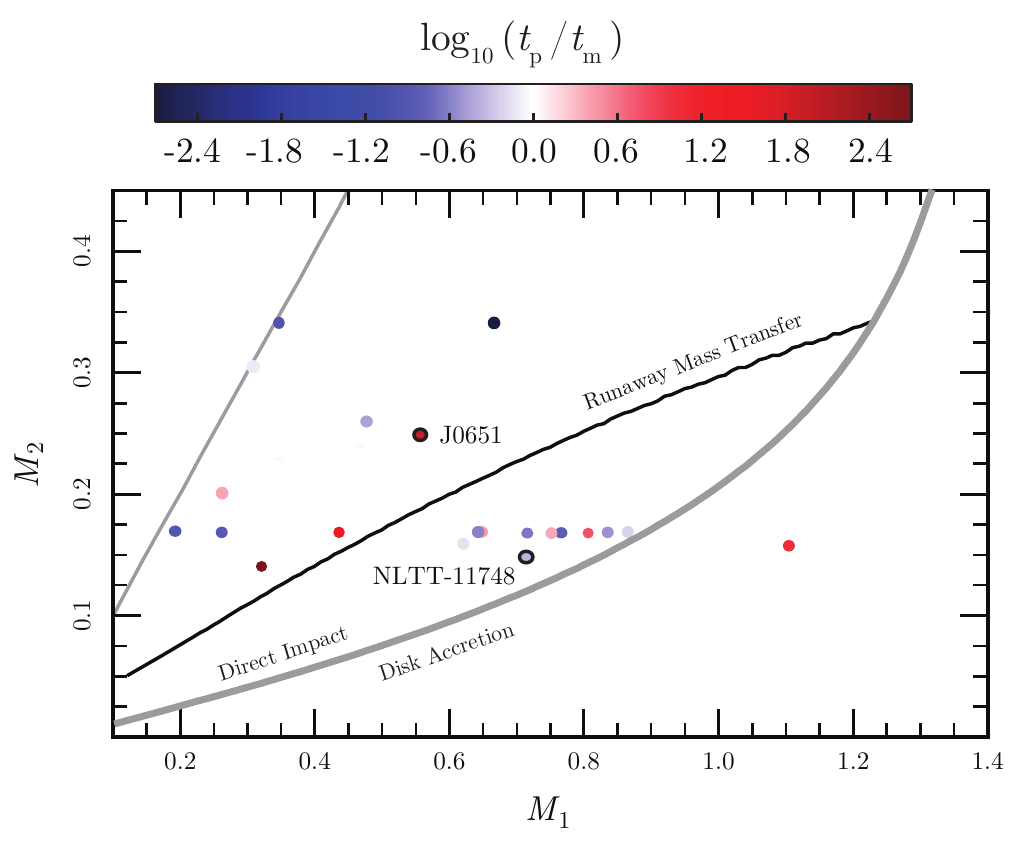} 
  \caption{The  \mone-\mtwo \,   diagram showing the observed DWD merger systems from \citet{Kilic2012} and including lines indicating the thresholds for disk formation and unstable mass transfer for a cold He WD \citep[adapted from][]{Gokhale2007,Dan2012}. 
 In color the ratio of the hydrogen burning lifetime to the merging time is shown. } 
\label{fig:m1m2}  
\vspace{1mm}

\end{figure}
 
\section{Evolution of Hydrogen Burning ELM WDs}
 \label{sec:evolution}
 \subsection{Methods}
 
ELM He WDs consist of an inert helium core and a tenuous outer layer. During the hydrogen  shell burning phase, the conductive interior is nearly isothermal with a temperature given by the hydrogen burning temperature of $T_{\rm core}\approx$ 10$^8$ K and evolution takes place at nearly constant luminosity dictated by the core mass - luminosity relation.

  \begin{figure*}[ht]
     \begin{center}
     
        \subfigure{%
            \label{fig:25_g_teff}
            \includegraphics[width=0.49\textwidth]{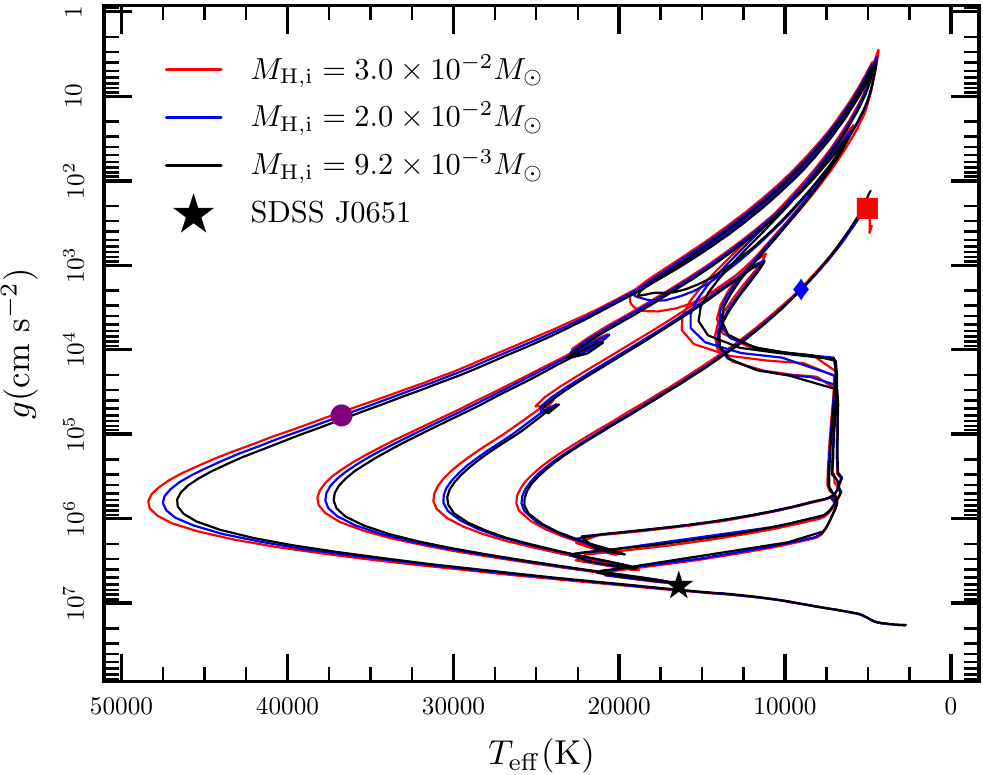}
        }
        \subfigure{%
           \label{fig:17_g_teff}
            \includegraphics[width=0.49\textwidth]{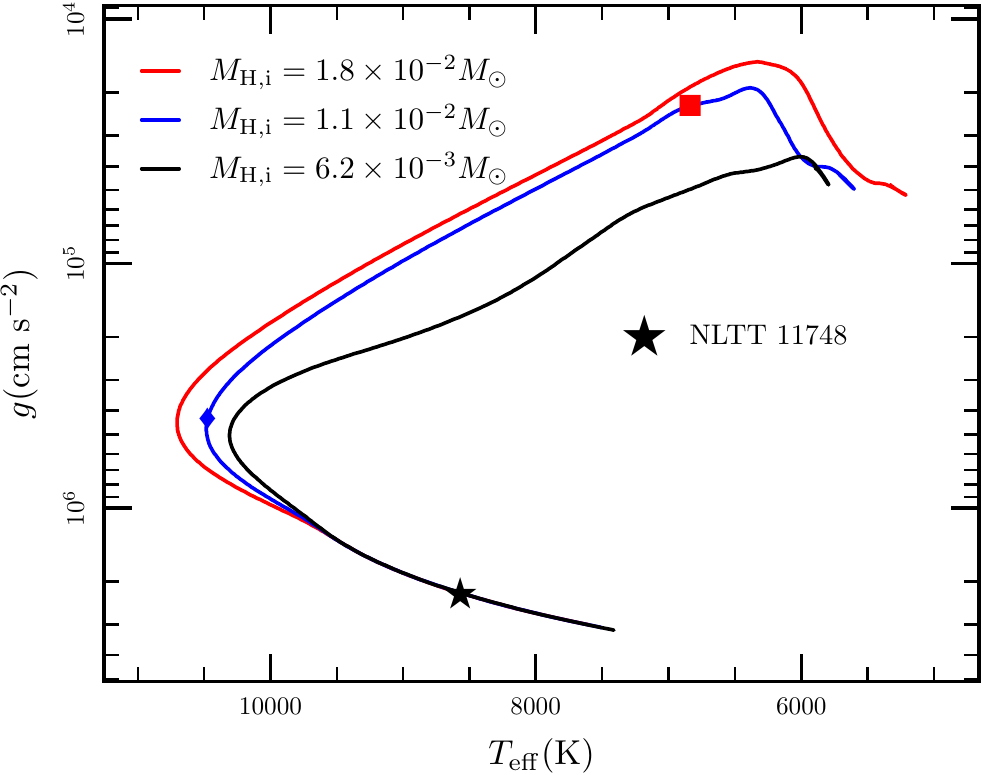}
        } 
    \end{center}
  \caption{The evolutionary tracks  of ELM helium WDs possessing hydrogen envelopes  with  \mtwo = 0.17 \msun and 0.25 \msun, representative of NLTT-11748 and J0651, respectively. Convergence is seen for the flashing models quickly along the cooling curve, where as the quiescent burning taking place on the lower mass models render them distinct. The red square, blue diamond, and purple circle show where the hydrogen mass is $10^{-2}, 5\times 10^{-3}$, and $ 10^{-3}$ \msun, respectively. The blue models are used for the rest of the figures.}
\label{fig:g_teff}  
\vspace{1mm}

\end{figure*}

To calculate the evolution of these ELM helium WDs  with hydrogen envelopes, we use Modules for Experiments in Stellar Astrophysics (MESA) \citep{Paxton2011, Paxton2013} version 6794. We model  helium WD formation by truncating a 1.0 \msun\,  star along the main sequence and removing mass. We then allow the inert core to come into diffusive equilibrium before allowing it to cool. 

As seen in other theoretical studies, ELM WDs within a mass range of about 0.18-0.36 \msun\, experience hydrogen CNO flashes due to a thermal instability induced by their geometrically thin burning shells \citep{Webbink1975, Althaus2001,Panei2007,Althaus2013,Gautschy2013}. The resulting runaway is quenched once the shell is no longer thin and leads to a rapid expansion of the stellar envelope. The WD may undergo several flashes until eventually the burning shell becomes too thin to maintain a sizable temperature perturbation and the WD burns the rest of its hydrogen stably. For WDs with masses either above or below the flashing range, the hydrogen content of the envelope is always burned quiescently until the WD joins the cooling sequence.  In order to calibrate our models, we focus on the two known eclipsing detached DWDs, which fortunately happen to sample both sides of this evolutionary hydrogen burning dichotomy.

SDSS J0651+2844 is the shortest period detached binary discovered, with an orbital period of $P_{\rm orb}$ = 12.75 min \citep{BrownJ0651} and provides an example of a system whose donor star lies within the flashing regime. The lower mass WD is found to have a mass of \mtwo = 0.25 \msun and a radius \rtwo = 0.0353$\pm$0.0004 \rsun, and the system is expected to come into contact  through gravitational waves in a time \tmerge = 0.9 Myr. 

A well studied example of a system that lies outside of the flashing regime, NLTT-11748, was discovered in 2009 as one of the first ELM WDs  with subsequent radial velocity observations revealing modulation about an orbital period of \porb = 5.64 h \citep{Kawka2009, Kawka2010}. In a search for radial pulsations within the lower mass WD, \citet{Steinfadt2010} discovered both primary and secondary eclipses within the system consistent with the measured orbital period. More recently, \citet{Kaplan2014} used ULTRACAM observations to  fit the eclipsing lightcurve. Their statistical analysis yields a WD mass of \mtwo = 0.17 $\pm$ 0.013 \msun\, and a radius of \rtwo = 0.0428 $\pm$ 0.005 \rsun. In addition to providing model independent masses and radii for the system, \citet{Kaplan2014} are able to measure the temperature of the companion CO WD 7600 $\pm$ 120 K and thus estimate a cooling age of 1.6--1.7 Gyr. \Fref{g_teff} shows the evolutionary tracks  of our models in the $\log g-T_{\rm eff}$ plane, which successfully  describe the  the properties of the observed systems.  

To calculate the past and future evolution due to gravitational radiation, we take the observed properties of the systems today (\mone, \mtwo, \porb) and time evolve the Roche lobe using the   equation for losses due to gravity waves 
\beqn
R_{\rm L1}(t) = (- \alpha t + R^4_{\rm L1,0})^{1/4}
\eeqn
where $\alpha = 256 G^3M_1 M_2 (M_1 + M_2)/(5c^5)$ and $R_{\rm L1,0}$ is the Roche lobe radius today, approximated analytically by \citet{Eggleton1983} as
\beqn 
R_{\rm L1,0} = \frac{0.49 q^{2/3}}{0.6 q^{2/3} + {\rm ln}(1+q^{1/3})} a_0
\eeqn  
where $a_0 = f$(\mone,\, \mtwo,\, \porb) is the observed separation today. 

Here we denote $t$ = 0 as the moment when our stellar model matches current observations of $\log g$ and $T_{\rm eff}$. We are then able to look at the past history and future evolution of our donor model in tandem with the analytically describable evolution of the Roche lobe.
Within this convention, {\it reattachment} takes place at a time $t  > 0$ when $R_{\rm L1}(t) = R_2(t)$ (from MESA) and CE {\it detachment} occurs at a time $ t < 0 $ when the same condition is satisfied. Since we are using the formula for point masses, this assumes that tidal interactions are not important until mass transfer occurs. 
 
 \subsection{Past Evolution: The Role of the CNO Flashes}
 \label{sec:past}
For WDs which undergo CNO flashes, we find that the past history of hydrogen burning is quickly eradicated. \Fref{25_mass_converge} shows the time evolution of hydrogen mass for two MESA WD models with the same total mass ($M_{\rm tot} = M_{\rm env} + M_{\rm core} $= 0.25 \msun) but very different initial envelope masses converging within less than 1 Gyr. Because the WDs have the same total mass, this means that the observables today would be independent of the initial conditions. 
\Fref{25_gw} shows the evolution of  $R_2$ and $R_{\rm L1}$   of a J0651-like 0.25 \msun\, WD undergoing CNO flashes. During the flashes $R_2>R_{\rm L1}$  and the ensuing mass transfer, which is not taken into account in our models, further prevents us from placing any constraints on the system's conditions at CE detachment.  
 
 \begin{figure}[t!]
     \begin{center}
     
        \subfigure{%
            \label{fig:25_mass_converge}
            \includegraphics[scale=0.75]{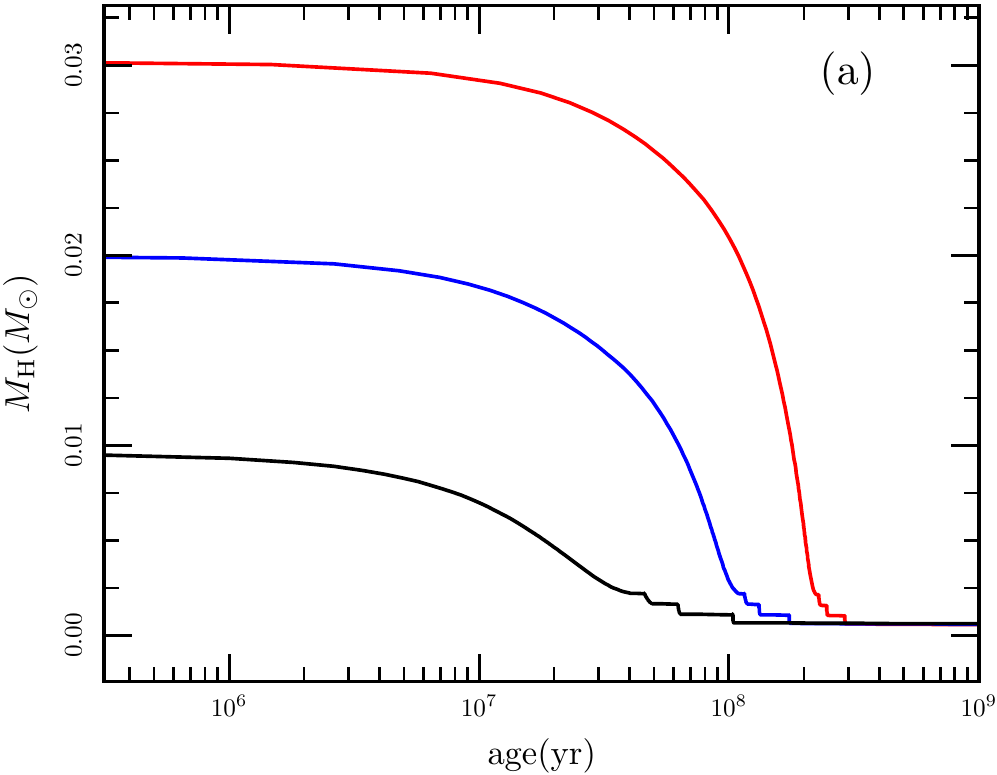}
        }\\
        \subfigure{%
           \label{fig:25_gw}
                        \includegraphics[scale=0.75]{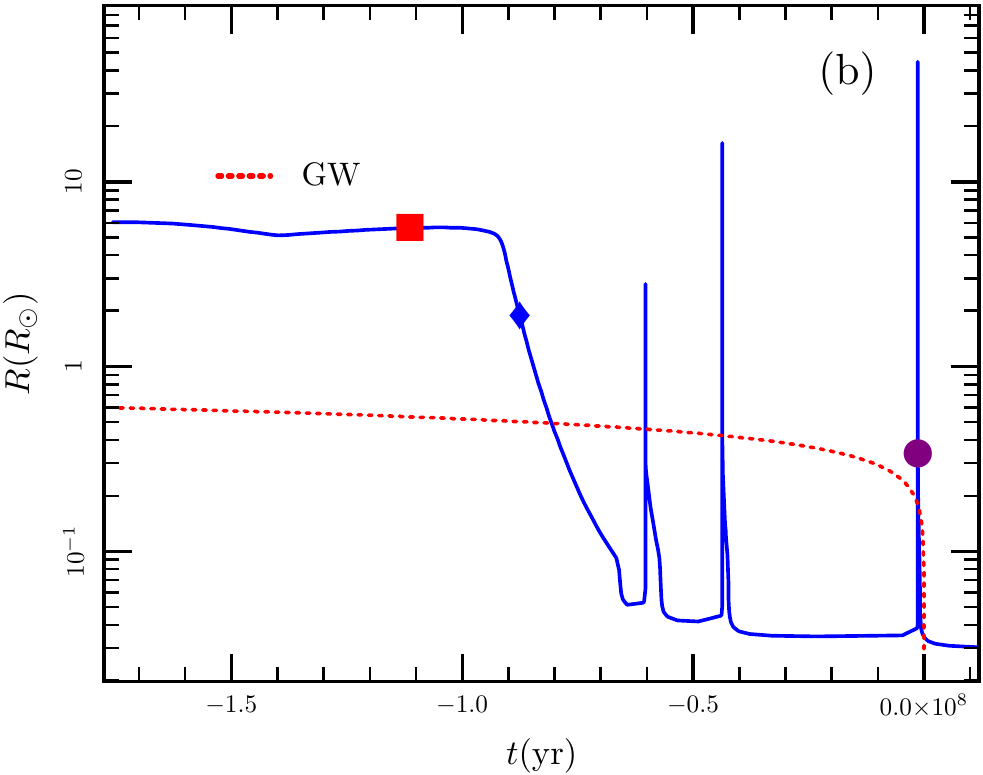}
        } 
    \end{center}
    \caption{%
\emph{Panel a}: Time evolution of total hydrogen mass for a 0.25 \msun WD with different initial hydrogen mass. The flashes quickly cause the two tracks to converge, thus eradicating their histories and demonstrating the inability of our method to constrain the past evolutionary history of systems with masses in the flashing range. \emph{Panel b}: Time evolution of both the WD radius due to hydrogen burning including the violent CNO flashes as well as $R_{\rm L1}$  due to gravitational waves. The x-axis is set such that t=0 corresponds to the observed radii today, whereas age does not include the offset. While we can see that the system will come come into contact quickly, the presence of the flashes does not allow us to uniquely trace back the radius evolution to detachment. }%
     
   \label{fig:25}
\end{figure}
 
For hydrogen burning ELM WDs outside of this flashing regime, the radius of the donor is  monotonically decreasing in time. \Fref{17_mass_converge} demonstrates the lack of convergence in time for different initial hydrogen masses since burning remains quiescent throughout  the evolution. This implies that an observation of WD mass and radius today allows us to uniquely constrain the  hydrogen mass and use our stellar models to map back to the CE detachment given some independent estimate of its post CE age. 

 \begin{figure}[t!]
     \begin{center}
     
        \subfigure{%
            \label{fig:17_mass_converge}
            \includegraphics[scale=0.75]{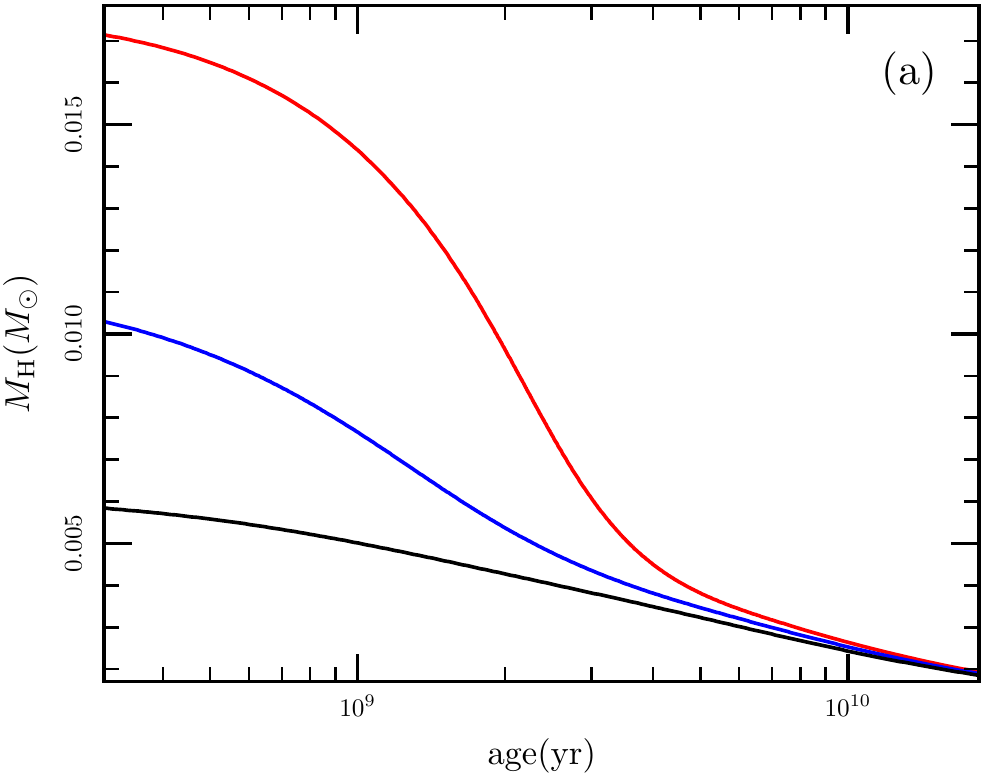}
        }\\
        \subfigure{%
           \label{fig:17_gw}
            \includegraphics[scale=0.76]{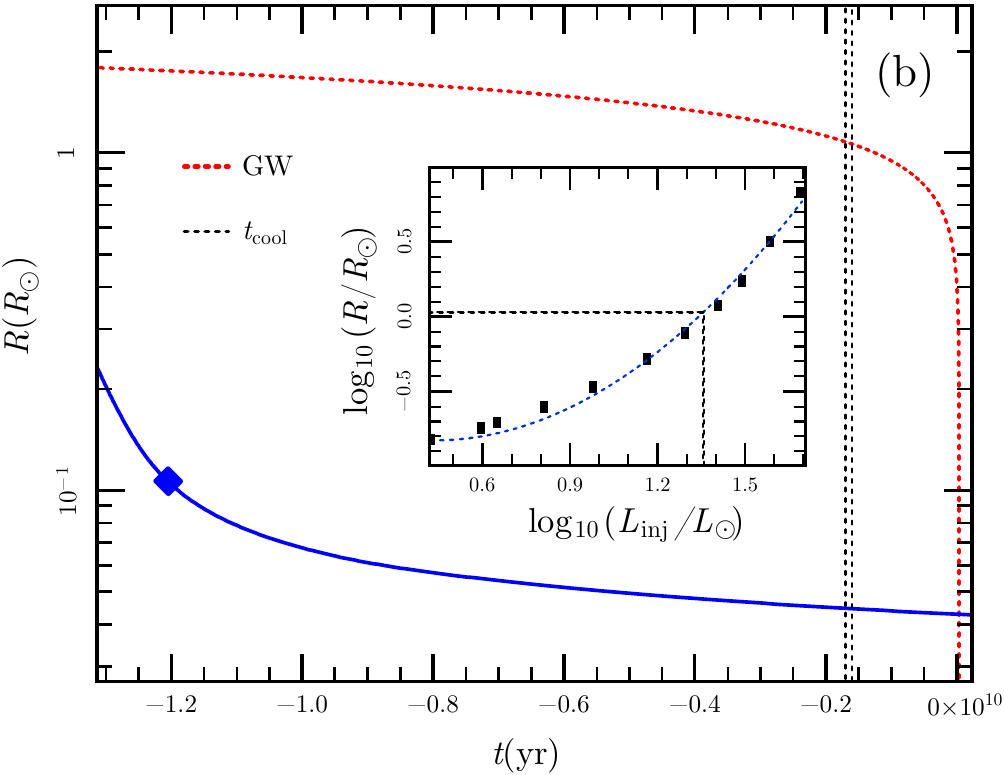}
        } 
    \end{center}
    \caption{% 
    {\it Panel a}: Time evolution of total hydrogen mass for a 0.17 \msun WD with different initial hydrogen mass. The WD burns it's hydrogen quiescently, thus allowing for the conditions today to self consistently be traced by to conditions at CE detachment.  \emph{Panel b}: Time evolution of the WD radius due to  hydrogen burning as well as \rlone . The blue curve represents the evolution due to quiescent hydrogen burning. In order have the system at in contact at a time which matches the cooling age of the CO WD companion, we find we need to inject a luminosity of $\sim$ 10$^{1.35}$ \lsun, which is shown by the dashed  blue curve in the inset panel. }%
     
   \label{fig:17}
\end{figure}

  \begin{figure*}[ht]
     \begin{center}
     
        \subfigure{%
            \label{fig:25_radial_comp}
            \includegraphics[width=0.47\textwidth]{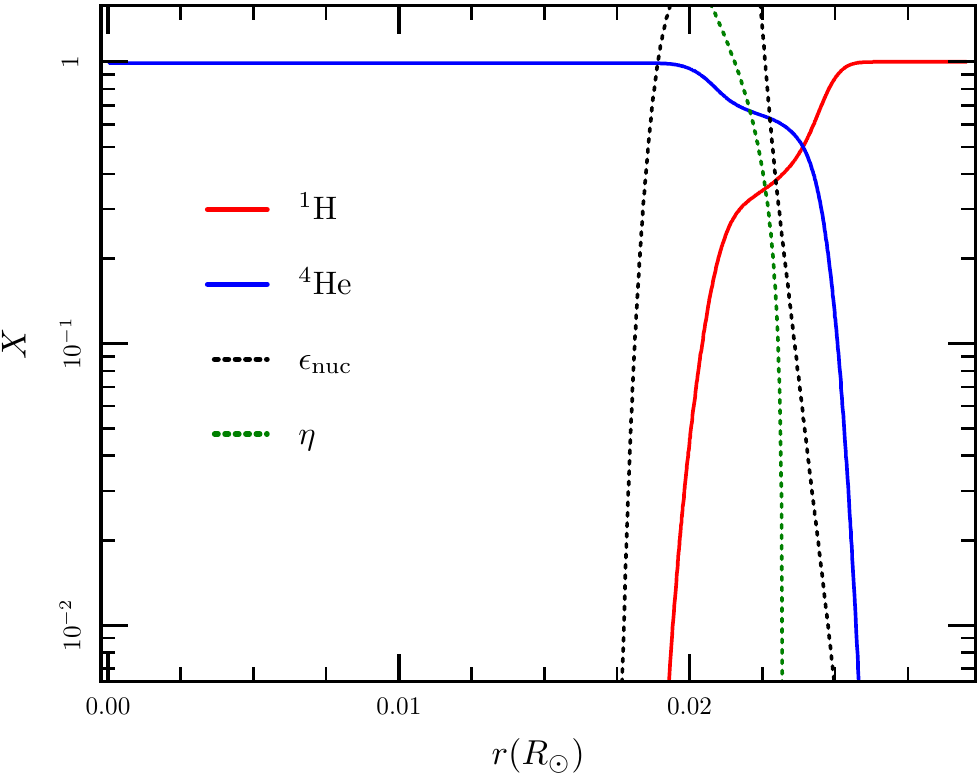}
        }
        \subfigure{%
           \label{fig:17_radial_comp}
            \includegraphics[width=0.47\textwidth]{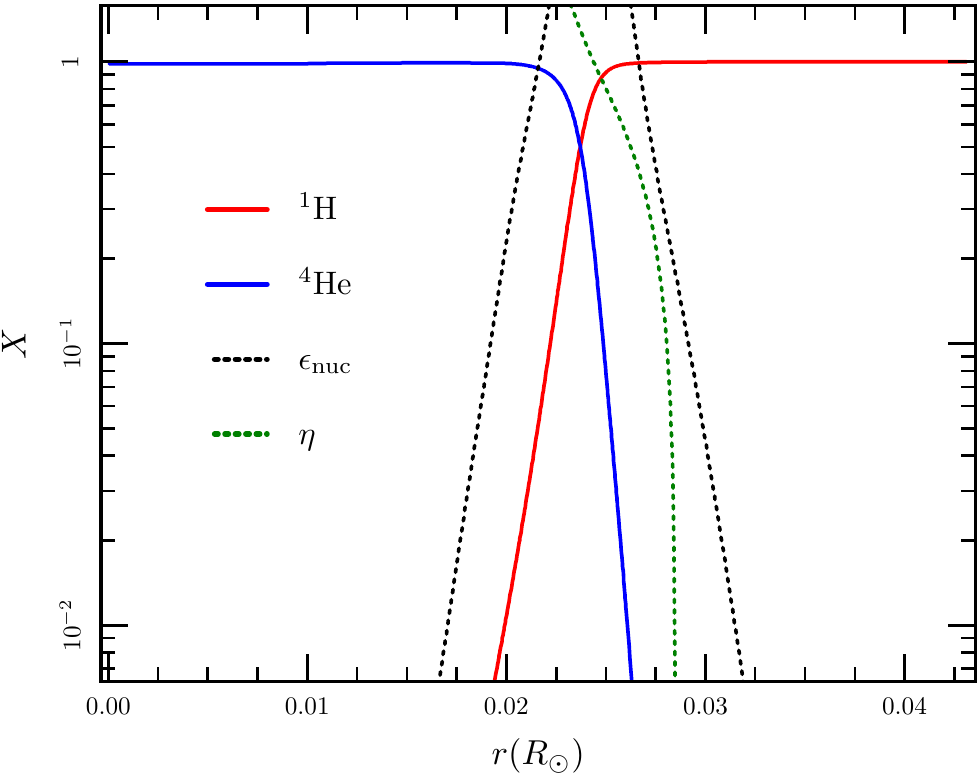}
        } 
    \end{center}
    \caption{%
Radial composition of the donor star in J0651 (\mtwo = 0.25 \msun) and NLTT-11748 (\mtwo = 0.17 \msun) at the time of reattachment ({\it left} and {\it right} panels, respectively). The black and green dotted lines show the specific energy generation rate and dimensionless degeneracy parameter $\eta$, respectively. In both systems, it is clear that the hydrogen burning envelope remains active until reattachment.  }%
     
   \label{fig:radial_comp}
\end{figure*}
Since the cooling properties of CO WDs are much better calibrated and do not depend on the possibility of CNO flashes and the cooling time of an insulating H burning envelope, they provide us with a reliable estimate of the time since the system emerged from a CE. With this, we trace the evolution of a 0.17 \msun\, WD with an initial hydrogen envelope mass of $1.1 \times 10^{-2}$ \msun\, and find that $M_{\rm H} (-t_{\rm cool}) = 2.2 \times 10^{-3}$ \msun.

Perhaps unsurprisingly, we find that we are unable to have our quiescently burning models in contact with the Roche lobe radius at the cooling age of the CO WD, regardless of the initial hydrogen mass. The solution to this ostensible paradox is to remember that during the CE phase there is a considerable injection of entropy onto the ELM WD, resulting in a substantial increase in radius. However, because of the relatively short thermal time of the envelope this entropy injection is almost immediately forgotten once the CE phase has ended. As a toy model, we modify MESA to artificially inject luminosity within the outer 10$^{-3}$ \msun\, of the WD in order to mimic the last phases of the CE in an NLTT-11748 like model and find that the CE phase must have provided a luminosity of  about $L_{\rm CE} \approx 8.6 \times 10^{34}$ erg s$^{-1}$ in order to have the WD in contact at the cooling age of the CO WD. Attributing this to an accretion luminosity, this corresponds to an accretion rate of  $\dot{M} = 5 \times 10^{-6}$ \msun\, yr$^{-1}$ (Figure \ref{fig:17_gw}).

\subsection{Future Evolution: Conditions at the Onset of RLO}
Because a hydrogen mass maps almost uniquely to a total stellar radius for a given total WD mass, we can use the observations of our two eclipsing systems today to infer  their future evolution. In particular, we wish to  investigate the thermal state of the donor at reattachment. 

\Fref{radial_comp} shows the composition, burning regions, and degeneracy parameter $\eta = E_{\rm F}/k_{\rm B} T$ as a function of radius for our model WDs at reattachment. At this stage, the 0.25 and 0.17 \msun\,  models have $6 \times 10^{-4} $ and $ 2 \times 10^{-3}$ \msun\, of hydrogen left, respectively, and are still undergoing hydrogen burning. Although the amount of mass in this burning envelope is minuscule, it results in a nearly 30\% increase in radius for both stars which leads them to begin to transfer mass at larger separations. This extended envelope is also significantly non-degenerate due to the heat released by the burning, leading to an inversion of the typical cold WD mass-radius relation. This inversion renders the system stable regardless of the mass ratio and would  likely result in an X-ray emitting, low frequency gravitational wave source.
 The duration of this phase of mass transfer depends sensitively on the interplay between gravitational wave losses, hydrogen burning lifetime of the donor, response of the accretor \citep{Shen2015}, and tidal torques which we plan to explore in a subsequent paper using the framework outlined in \citet{Gokhale2007}.

 \section{Discussion}
 \label{sec:discussion}
 In this {\it Letter}, we have demonstrated a method of combining stellar evolution models in tandem with standard GW calculations to infer the past and future evolution of currently detached DWD binaries. Using this method we are for the first time able to put constraints on donor conditions such as envelope mass and luminosity immediately proceeding the CE given the current separation, masses, and donor radii of these detached DWDs, assuming they are not within the flashing regime and have a reliable cooling age for the higher mass WD. Regardless of the presence of flashes, we are able to say whether or not the donor will be actively burning hydrogen at the onset of mass transfer and thus whether standard stability calculations can predict if the system will truly merge at contact. 
  
 We find that \emph{roughly half} of the currently observed detached DWDs may begin mass transferring when the donor is still burning hydrogen, resulting in larger separations at the onset of mass transfer, lower accretion rates, and a period derivative consistent with evolution driven by gravitational radiation. Self-consistently modeling of  the resulting mass transfer for these systems  is required to infer  the lifetime of the non-degenerate, hydrogen-rich mass transfer  phase, which in turn is essential for predicting the number of X-ray and low frequency gravitational wave sources. The properties of the surviving population depend critically on the interplay of these physical processes and can be used to constrain them. To this end, a detailed study  of the stabilizing effect of tidal interactions during this mass transfer phase is needed, as the current $10^6$ yr lifetime predicted  by \citet{D'Antona2006} and \citet{Kaplan2012} appears to be too short to explain the presence of HM Cancri and V407 Vulpeculae.  
 
While the current number of DWDs in which the donor radius is known is small due to the necessity of an eclipse, {\it Gaia} will greatly increase this sample by accurately determining the distance to many systems \citep{Holberg2012}.  By applying  this technique to a population of  DWDs with accurate donor radii one  can  statistically constrain the conditions at  CE  detachment, enabling statistical studies that can help gain insight into  the formation of compact binaries.  \\

We thank J. Guillochon, M. Macleod, and T. Tauris for useful discussions. We acknowledge support from the David and Lucile Packard Foundation, Radcliffe Institute for Advanced Study, and the National Science Foundation Graduate Research Fellowship.

 \end{document}